# A new stochastic differential equation approach for waves in a random medium

Dimitris Dimitropoulos and Bahram Jalali

Optoelectronic Circuits and Systems Lab, University of California, Los Angeles,

CA90095

**Abstract** – We present a mathematical approach that simplifies the theoretical treatment of electromagnetic localization in random media and leads to closed form analytical solutions. Starting with the assumption that the dielectric permittivity of the medium has delta-correlated spatial fluctuations, and using Ito's lemma, we derive a linear stochastic differential equation for a one dimensional random medium. The equation leads to localized wave solutions. The localized wave solutions have a localization length that scales as $L \sim \omega^{-2}$ for low frequencies whereas in the high frequency regime this length behaves as $L \sim \omega^{-2/3}$.

Since the original prediction by Anderson [1] that a random medium supports localized wave solutions an intense interest has been generated in observing and predicting the behaviour of these waves. The effect is difficult to observe in a solid state electronic system, however, it has been observed with electromagnetic waves[2]. Both weak and strong localization of light have been observed [3,4,5] supported by theoretical work that predict their properties [6,7,8]. Independent of the theoretical approach taken, the mathematical treatments of wave propagation in random media tend to be complicated and in most cases evade a closed form solution. To obtain the relevant behavior, the analytical treatments must be followed by numerical simulations.

In this paper, we report a theoretical approach to electromagnetic localization in random media that is surprisingly simple. We show that by using a modified form of Ito's lemma [9], a stochastic calculus technique popular in the world of mathematical finance, we can obtain a stochastic differential equation for waves in a one-dimensional random medium that readily leads to closed form analytical solutions. To the best of our knowledge, only other stochastic differential equation approach we are aware of is for the Schrodinger equation by Dawson et al [10]. Our approach allows the permittivity of the material to attain negative values and therefore model a certain "metallic" component in the medium. Our approach leads to localized waves with localization length that scales with frequency as $L \sim \omega^{-2}$, at low frequency, a behavior well established by other approaches [11,12]. In the high-frequency regime, the localization length scales as $L \sim \omega^{-2/3}$.

We begin by considering the electric and magnetic field equations in a one-dimensional medium with inhomogenous dielectric permittivity:

$$\frac{\partial}{\partial x} E = -\frac{\partial}{\partial t}(\mu H) \qquad (1a)$$

$$\frac{\partial}{\partial x} H = -\frac{\partial}{\partial t}\left((\varepsilon + \Delta\varepsilon(x))E\right) \qquad (1b)$$

where $E, H$ are the electric and magnetic field and $\mu, \varepsilon$ the magnetic and electric permittivities and $\Delta\varepsilon(x) = \varepsilon \cdot \sigma \cdot \zeta(x)$ is the random permittivity perturbation. The parameter $\sigma$ dictates the magnitude of the random spatial variations of the permittivity. For simplicity, the medium is assumed to be loss-less although approach is capable of taking losses into account. The usual effect of losses is that they make localization more difficult to observe [2]. We would like to solve the above equations for a random function $\zeta(x)$ that is delta correlated in space:

$$\langle \zeta(x)\zeta(x') \rangle = \delta(x-x') \tag{2}$$

The function $\zeta(x)$ (units are $cm^{-1/2}$ since the delta function has units $cm^{-1}$) describes fluctuation in the permittivity and can obtain both positive and negative values. This correlation function represents zero correlation length and allows us to invoke Itoh's lemma. The resulting delta-correlated permittivity is an approximation that is valid when the solution varies slowly over distances of the correlation length $a$. When the spatial variation of the solution approaches the correlation length our solutions are no longer appropriate as the exact functional form of the correlation function becomes important at these length scales.

Integrating equations (1a) and (1b) we obtain:

$$E(x,t) - E(x',t) = -\mu \int_{x'}^{x} dx'' \frac{\partial H}{\partial t} \tag{3a}$$

$$H(x,t) - H(x',t) = -\varepsilon \int_{x'}^{x} dx'' \frac{\partial E}{\partial t} - \sigma\varepsilon \int_{x'}^{x} \frac{\partial E}{\partial t} \zeta(x'') dx'' \tag{3b}$$

We now define a new stochastic variable, $Z(x) = \int_{0}^{x} \zeta(x')dx'$ (has units $cm^{1/2}$), that captures the random permittivity in equation 3b, and consider the dependence of electric and magnetic fields on it: $E = E(x,t,Z(x))$ and $H = H(x,t,Z(x))$. We assume that the higher moments of the random variable $\zeta(x)$ are such that the function $dZ(x) = \zeta(x)dx$ is a Gaussian distributed variable with $E(dZ^2) = dx$ and therefore $Z(x)$ is a Brownian motion random variable [9]. In this case the differentials of the functions $E, H$ can be expressed using Ito's lemma (see Appendix):

$$dE = \frac{\partial E}{\partial t} dt + \left( \frac{\partial E}{\partial x} + \frac{1}{2} \frac{\partial^2 E}{\partial Z^2(x)} \right) dx + \frac{\partial E}{\partial Z(x)} dZ(x) \qquad (4a)$$

$$dH = \frac{\partial H}{\partial t} dt + \left( \frac{\partial H}{\partial x} + \frac{1}{2} \frac{\partial^2 H}{\partial Z^2(x)} \right) dx + \frac{\partial H}{\partial Z(x)} dZ(x) \qquad (4b)$$

The reason the 2$^{nd}$ order derivative appears in the linear expansion stems from the fact that $E(dZ^2) = dx$ and therefore 2$^{nd}$ order changes in the variable $Z(x)$ converge to a 1$^{st}$ order change in the variable $x$. A term by term comparison of equations (4a) and (4b) with equations (3a) and (3b) gives the following four equations:

$$\frac{\partial E}{\partial x} = -\mu \frac{\partial H}{\partial t} \qquad (5a)$$

$$\frac{\partial E}{\partial Z(x)} = 0 \qquad (5b)$$

$$\frac{\partial H}{\partial x} + \frac{1}{2} \frac{\partial^2 H}{\partial Z^2(x)} = -\varepsilon \frac{\partial E}{\partial t} \qquad (5c)$$

$$\frac{\partial H}{\partial Z(x)} = -\sigma \varepsilon \frac{\partial E}{\partial t} \qquad (5d)$$

Combining the equations (5a-d) we obtain two equations for the magnetic field:

$$\frac{\partial H}{\partial x} + \frac{1}{2} \frac{\partial^2 H}{\partial Z^2(x)} = \frac{1}{\sigma} \frac{\partial H}{\partial Z(x)} \qquad (6a)$$

$$\frac{\partial^2 H}{\partial x \partial Z(x)} = \sigma \mu \varepsilon \frac{\partial^2 H}{\partial t^2} \qquad (6b)$$

Once the magnetic field is calculated the electric field is then obtained through (3a). Substituting the solution $H(t, x, Z(x)) = \tilde{H}(\omega, k, \lambda) \exp(-j\omega t + jkx + j\lambda Z(x))$ in (6a) and (6b) we obtain:

$$jk - (1/2)\lambda^2 = j(\lambda/\sigma) \tag{7a}$$

$$k\lambda = \omega^2 \mu\varepsilon\sigma \tag{7b}$$

With the substitution $k/k_o = jA$ (where $k_o^2 = \omega^2\mu\varepsilon$) we can obtain dimensionless equations where the dimensionless parameter $k_o\sigma^2$ controls the solution:

$$A^3 + A - (1/2)k_o\sigma^2 = 0 \tag{8a}$$

$$\frac{\lambda}{\sigma} = \frac{k_o}{jA} \tag{8b}$$

If we consider the expectation value for the wave amplitude we get (for $x > 0$):

$$\langle \exp(-j\omega t + jkx + j\lambda Z(x)) \rangle = \exp(-j\omega t + j(\lambda/\sigma)x) \tag{9}$$

where we make use of the expectation value $\langle \exp(j\lambda Z(x)) \rangle = \exp(-(1/2)\lambda^2 x)$ and of equation (7a).

When $(1/2)k_o\sigma^2 \gg 1$ equation (8a) has approximate solutions, $A \cong \exp(j120°)(k_o\sigma^2/2)^{1/3}$ and $A \cong \exp(-j120°)(k_o\sigma^2/2)^{1/3}$, or $A \cong (k_o\sigma^2/2)^{1/3}$. Let us consider the implications of the first two solutions. These along with equation (9) result in a solution [13]:

$$H \propto \exp\left(-\frac{k_o x}{2|A|}\right)\cos\left(\omega t - \frac{\sqrt{3}k_o x}{2|A|}\right) \tag{10}$$

This solution has the general form of a localized wave: $\exp(-x/L)\cdot\cos(\omega t - k_{eff}x)$, with a particular dispersion relationship $k_{eff}(\omega) \sim \omega^{2/3}$ and a localization length $L \sim \omega^{-2/3}$.

The third and real solution for $A$ ($A \cong (k_o \sigma^2 / 2)^{1/3}$) corresponds to an exponentially increasing solution $H \propto \exp\left(\dfrac{k_o x}{|A|}\right)$. This is an un-physical solution since there is no energy source in the problem that can make the wave amplitude grow and we therefore reject it.

Similarly if we consider the limit $(1/2)k_o \sigma^2 \ll 1$ the approximate solutions of (8a) are $A \cong k_o \sigma^2 / 2$, $A \cong \pm j - k_o \sigma^2 / 4$. The two complex solutions give the localized waves in that limit, with a localization length $L \cong 4/(k_o^2 \sigma^2) \sim \omega^{-2}$ and an effective wavevector $k_{eff}(\omega) \cong k_o$.

In the figure below we show the localization length and the "effective wavelength" $\lambda_{eff} = 2\pi / k_{eff}$ for the localized wave solution. The variation with respect to the dimensionless parameter $(1/2)k_o \sigma^2$ is plotted and all quantities are normalized to the material wavelength $\lambda_o = (2\pi / k_o)$.

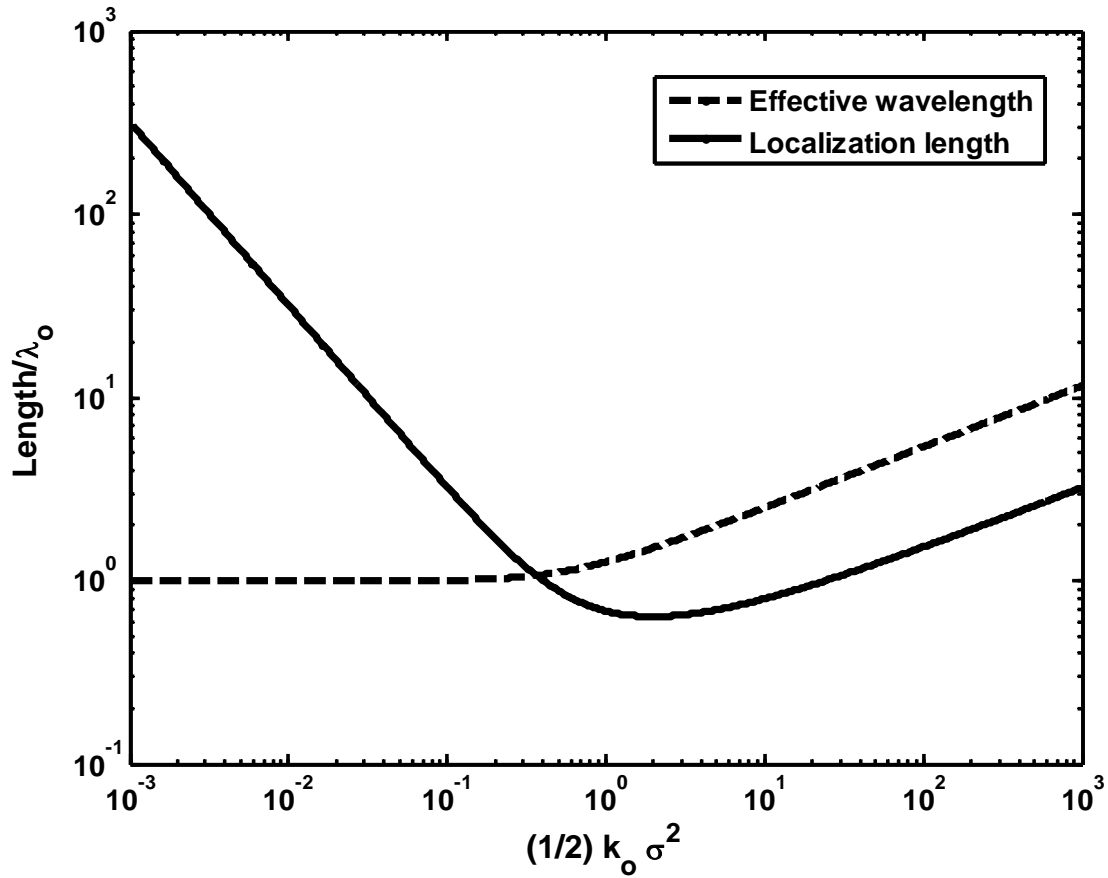

**Figure 1: The localization length and the effective wavelength for the wave solution is shown. All quantities are normalized to the material wavelength.**

The localized wave solution has a localization length that scales with frequency as $L \sim \omega^{-2/3}$ at high frequencies and as $L \sim \omega^{-2}$ at low frequencies. The "effective" wavelength does not exhibit much of a change in the low frequency regime, in contrast with the behavior of the localization length. In the high-frequency regime the effective wavelength and the localization length have a similar behavior.

In summary, we have obtained a stochastic differential equation for electromagnetic waves propagating in random medium with delta-correlated spatial fluctuations. In practice, materials have fluctuations of a finite correlation length so our solution will model physical problems accurately when it varies very slowly over length scales equal to the correlation length of the problem. The solutions to our equation are localized waves and the localization length scales as $L \sim \omega^{-2}$ for low frequencies and as $L \sim \omega^{-2/3}$ in the high frequency regime.

Appendix

The Ito lemma gives the differential of a function of stochastic variables of the form

$$dy = a(y,x)dx + b(y,x)dZ(x) \tag{A1}$$

where $dZ(x) \propto dx^{1/2}$ is a random process that follows Brownian motion. For any continuously differentiable function $f(y,t)$ [9] for which the 2nd derivative exists:

$$df = b\frac{\partial f}{\partial y}dZ(x) + \left(\frac{\partial f}{\partial x} + a\frac{\partial f}{\partial y} + \frac{b^2}{2}\frac{\partial^2 f}{\partial y^2}\right)dx \tag{A2}$$

To obtain equations (4a-b) use this relation with $a=0, b=1$, a condition that leads to: $dy = dZ$.

References

[1] P.W. Anderson, Phys. Rev. **109**, 1492 (1958)

[2] A.Z. Genack *et al* , J. Opt. Soc. Am. B **10**, 408 (1993)

[3] M.P. Van Albada *et al* , Phys. Rev. Lett. **55**, 2692 (1985)

[4] P-E. Wolf *et al* , Phys. Rev. Lett. **55**, 2696 (1985)